\documentclass[a4paper,twocolumn,superscriptaddress,showpacs,preprintnumbers,amsmath,amssymb,aps,pra,floatfix]{revtex4}
\usepackage{graphicx}	
\usepackage{subfigure}	
\usepackage{dcolumn}	
\usepackage{bm}		
\usepackage{hyperref}	

\usepackage[applemac]{inputenc}
\usepackage[T1]{fontenc}
\usepackage[british]{babel}
\usepackage{color}
\usepackage{upgreek}

\newcommand{\fbr}{Feshbach resonance}
\newcommand{\tof}{time-of-flight}
\newcommand{\iqoqi}{Institut f\"ur Quantenoptik und Quanteninformation}
\newcommand{\oeaw}{\"Osterreichische Akademie der Wissenschaften}
\newcommand{\ibk}{Innsbruck}
\newcommand{\1}{$|1\rangle$}
\newcommand{\2}{$|2\rangle$}
\newcommand{\3}{$|3\rangle$}
\newcommand{\lithium}{$^6$Li}
\newcommand{\kalium}{$^{40}$K}

\begin{document}

\title{All-optical production of a degenerate mixture of \lithium\ and \kalium\\
and creation of heteronuclear molecules}

\author{F.M. Spiegelhalder}
\author{A. Trenkwalder}
\author{D. Naik}
\author{G. Kerner}
\affiliation{\iqoqi, \oeaw, 6020 \ibk, Austria}
\author{E. Wille}
\altaffiliation[Current address: ]{European Space Agency, Noordwijk, The Netherlands}
\affiliation{\iqoqi, \oeaw, 6020 \ibk, Austria}
\affiliation{Institut f\"ur Experimentalphysik und Zentrum f\"ur Quantenphysik, Universit\"at \ibk, 6020 \ibk, Austria}
\author{G. Hendl}
\author{F. Schreck} \affiliation{\iqoqi, \oeaw, 6020 \ibk, Austria}
\author{R. Grimm}
\affiliation{\iqoqi, \oeaw, 6020 \ibk, Austria}
\affiliation{Institut f\"ur Experimentalphysik und Zentrum f\"ur Quantenphysik, Universit\"at \ibk, 6020 \ibk, Austria}

\date{\today}


\begin{abstract}
We present the essential experimental steps of our all-optical approach to prepare a double-degenerate Fermi-Fermi mixture of \lithium\ and \kalium\ atoms, which then serves as a starting point for molecule formation. We first describe the optimized trap loading procedures, the internal-state preparation of the sample, and the combined evaporative and sympathetic cooling process. We then discuss the preparation of the sample near an interspecies \fbr, and we demonstrate the formation of heteronuclear molecules by a magnetic field ramp across the resonance.
\end{abstract}


\maketitle

\section{Introduction}

The groundbreaking achievements in experiments with ultracold Fermi gases \cite{Inguscio2006ufg, Giorgini2008tou} have opened up unprecedented possibilities to study new regimes of strongly interacting quantum matter. Ultracold gases represent well-controllable model systems for the exploration of many-body regimes in a way not possible in conventional condensed-matter systems~\cite{Bloch2008mbp}. A new frontier in the field is currently being explored in experiments on ultracold Fermi-Fermi mixtures of \lithium\ and \kalium\ atoms \cite{Taglieber2008qdt, Wille2008eau, Voigt2009uhf, Spiegelhalder2009cso, Tiecke2009sfb}. Because of the mass imbalance and the possibility to apply species-specific optical potentials, such systems promise manifold intriguing applications both in many-body physics \cite{Paananen2006pia, Iskin2006tsf, Iskin2007sai, Petrov2007cpo, Iskin2008tif, Baranov2008spb, Bausmerth2009ccl, Nishida2009ipw, Wang2009qpd, Mora2009gso} and few-body physics \cite{Petrov2005dmi, Nishida2009cie, Levinsen2009ads}.



To prepare degenerate Fermi gases, all-optical approaches have proven to be simple and robust and they facilitate highly efficient evaporative cooling. Therefore they are routinely applied in many present experiments; see Ref.~\cite{Inguscio2006ufg} for a review of earlier work and Refs.~\cite{Fuchs2007mbe, Inada2008cta, Ottenstein2008cso, Huckans2009tbr} for more recent examples. Spin mixtures of $^6$Li atoms near a broad Feshbach resonance are particularly well suited for this cooling approach because of their exceptional collision properties, which offer extremely large cross sections for elastic collisions in combination with very weak inelastic decay. This favorable situation motivates the general idea of using the strongly interacting \lithium\ gas as a cooling agent for sympathetic cooling of another species.
Following this idea in Ref.~\cite{Spiegelhalder2009cso}, we recently demonstrated the sympathetic cooling of \kalium\ atoms by an evaporatively cooled, optically trapped spin mixture of \lithium, reaching the double-degenerate regime.

In this Article, we first present more details on our all-optical approach of preparing a double-degenerate Fermi-Fermi mixture of \lithium\ and \kalium. We then show new results related to interactions and molecule formation near interspecies \fbr s. In Sec.~\ref{sec:trapping}, we discuss our dual-species cooling and trapping setup and the special loading procedures used for the optical traps. In Sec.~\ref{sec:SpinRelax}, we present an important preparation step where spin relaxation in the mixture brings the K atoms into their lowest internal state. In Sec.~\ref{sec:Evaporation}, we describe the combined evaporative and sympathetic cooling process. In Sec.~\ref{sec:StatePrep}, we show how the mixture can be prepared near interspecies \fbr s. In Sec.~\ref{sec:Molecules}, we finally demonstrate the creation of ultracold heteronuclear Fermi-Fermi molecules by Feshbach association methods.


\section{Dual-species cooling and trapping setup and procedures}
\label{sec:trapping}

\begin{figure}[tb]
\begin{center}
\includegraphics[width=\columnwidth]{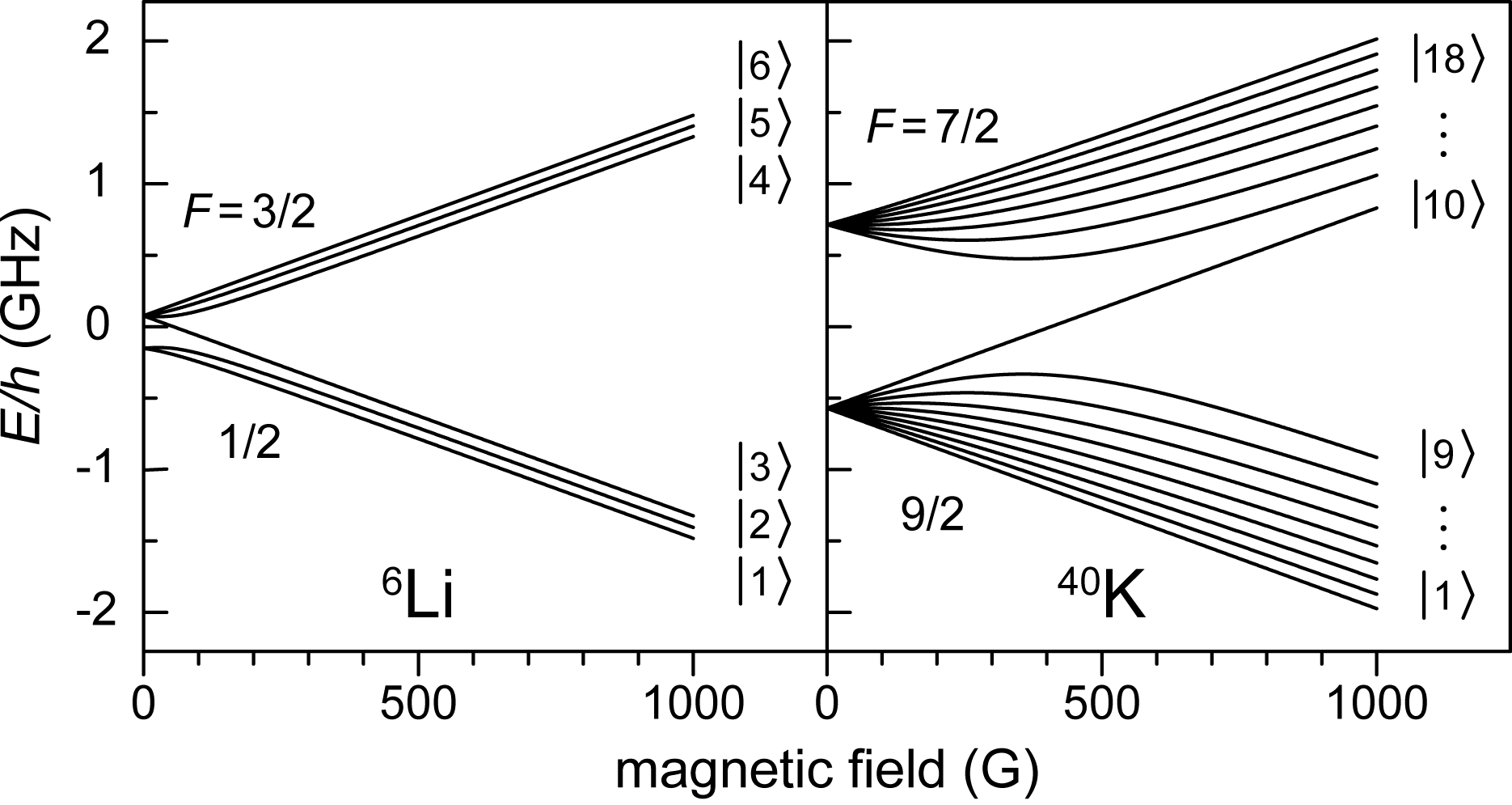}
\caption{Electronic ground state energies of \lithium\ and \kalium\ versus magnetic field.}
\label{fig:LevelScheme}
\end{center}
\end{figure}

Here, we outline the basic concept of our dual-species setup (Sec.~\ref{sec:setup}), and we present the special procedures applied to prepare the optically trapped mixture. In Sec.~\ref{sec:MOT} we describe how we operate a dual-species magneto-optical trap (MOT). In Sec.~\ref{sec:ODT} we present the optical dipole traps (ODT) used in the experiments. In Sec.~\ref{sec:transfer} we discuss the special loading procedure for the ODT. The whole scheme is optimized for a large number of \lithium\ atoms, as this species is used as the cooling agent for sympathetic cooling of \kalium\ into degeneracy \cite{Spiegelhalder2009cso}.

Figure~\ref{fig:LevelScheme} shows the atomic ground state energy structure of \lithium\ and \kalium. We label the energy levels Li$|i\rangle$ and K$|j\rangle$, counting the states with rising energy. The hyperfine splitting of \lithium\ is 228.2\,MHz. For \kalium, the hyperfine structure is inverted and the splitting amounts to 1285.8\,MHz \cite{Arimondo1977edo}. For the low-lying states with $i\leq3$ and $j\leq10$, the projection quantum numbers are given by $m_\mathrm{Li}=-i+3/2$ and $m_\mathrm{K}=j-11/2$.

\subsection{Experimental setup}
\label{sec:setup}

For the cooling and trapping of Li and K we apply standard laser cooling and trapping techniques \cite{Metcalf1999lca} combining a Zeeman-slowed atomic beam and a dual-species MOT for initial collection of atoms in the vacuum chamber. A detailed description of the experimental setup and the laser systems can be found in Ref.~\cite{Wille2009poa}. A dual-species oven, which is connected to the main chamber via a differential pumping section, delivers a well-collimated atomic beam. We operate the oven with isotopically enriched samples containing 80\% of \lithium\ and 7\% of \kalium. The Zeeman slower can cool both species individually with the respective settings of the magnetic field gradients. The central element of our vacuum chamber is a glass cell that allows for very good optical access. We achieve excellent vacuum conditions with a pressure on the order of 10$^{-11}$\,mbar.

\begin{figure}[tb]
\begin{center}
\includegraphics[width=\columnwidth]{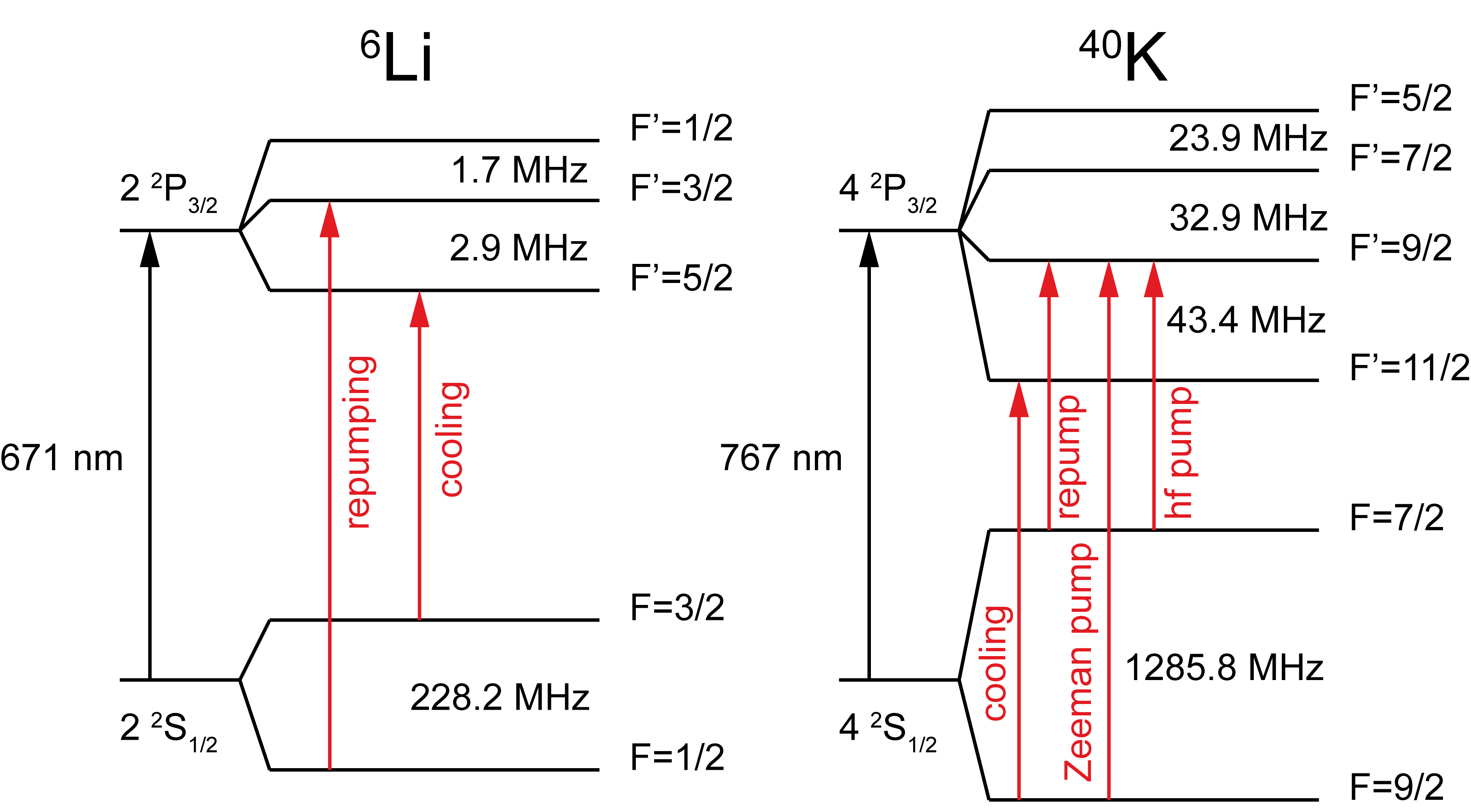}
\caption{Scheme of the atomic energy levels and transitions used to cool and trap Li and K. Also shown are the transitions used for hyperfine (hf) and Zeeman pumping of K.}
\label{fig:OptPump}
\end{center}
\end{figure}

For both species we use diode laser systems with one grating stabilized master oscillator in combination with tapered amplifiers. The Li (K) laser system provides 11\,mW (12\,mW) per MOT beam and 80\,mW (100\,mW) for the Zeeman slower beam. Figure~\ref{fig:OptPump} shows a schematic drawing of the atomic energy levels and optical transitions used for cooling and trapping of Li and K. The Li MOT laser beams contain two frequency parts tuned to the cooling ($F=3/2\rightarrow F'=5/2$) and repumping ($F=1/2\rightarrow F'=3/2$) transitions and having equal power. For K the cooling ($F=9/2\rightarrow F'=11/2$) to repumper ($F=7/2\rightarrow F'=9/2$) ratio is three to two.

\subsection{MOT loading}
\label{sec:MOT}

The initial collection and cooling of \lithium\ and \kalium\ is achieved by conventional laser-cooling and trapping techniques. As loading of both species requires different settings of the Zeeman slower magnetic field, we use a sequential MOT-loading scheme. The basic idea is to first load Li and then add K in the presence of the Li MOT.

The Li MOT is operated with a field gradient of 26\,G/cm along the symmetry axis of the field coils and a laser detuning of $-27$\,MHz, for both, cooling and repumping transition. After a loading time of about 15\,s, a few $10^9$ Li atoms are accumulated in the MOT. At this point we increase the magnetic field gradient to 38\,G/cm, where the K MOT works optimally. In 5\,s, about $10^7$ K atoms are added to the trap. The K MOT is operated with a laser detuning of $-34$\,MHz.

During the K loading phase we operate the Li MOT with a relatively large detuning of $-31$\,MHz in order to compensate for the effect of the higher magnetic field gradient on volume and density. This avoids increased losses by inelastic interspecies collisions, enabling the efficient sequential loading of both species.

In order to reduce shot-to-shot fluctuations of the number of atoms in the trap, we control the Li and K MOT loading by monitoring their fluorescence independently. When the fluorescence of the Li MOT reaches its threshold value, Li loading is stopped and loading of the K MOT is initiated. Once the K MOT fluorescence reaches its threshold, the K loading is turned off. At this point the ODT is ramped on in 100\,ms, and the magnetic fields of the Zeeman slower are ramped to zero in 10\,ms.

\subsection{Optical dipole trapping schemes}
\label{sec:ODT}

\begin{figure}[t]
\begin{center}
\includegraphics[width=0.7\columnwidth]{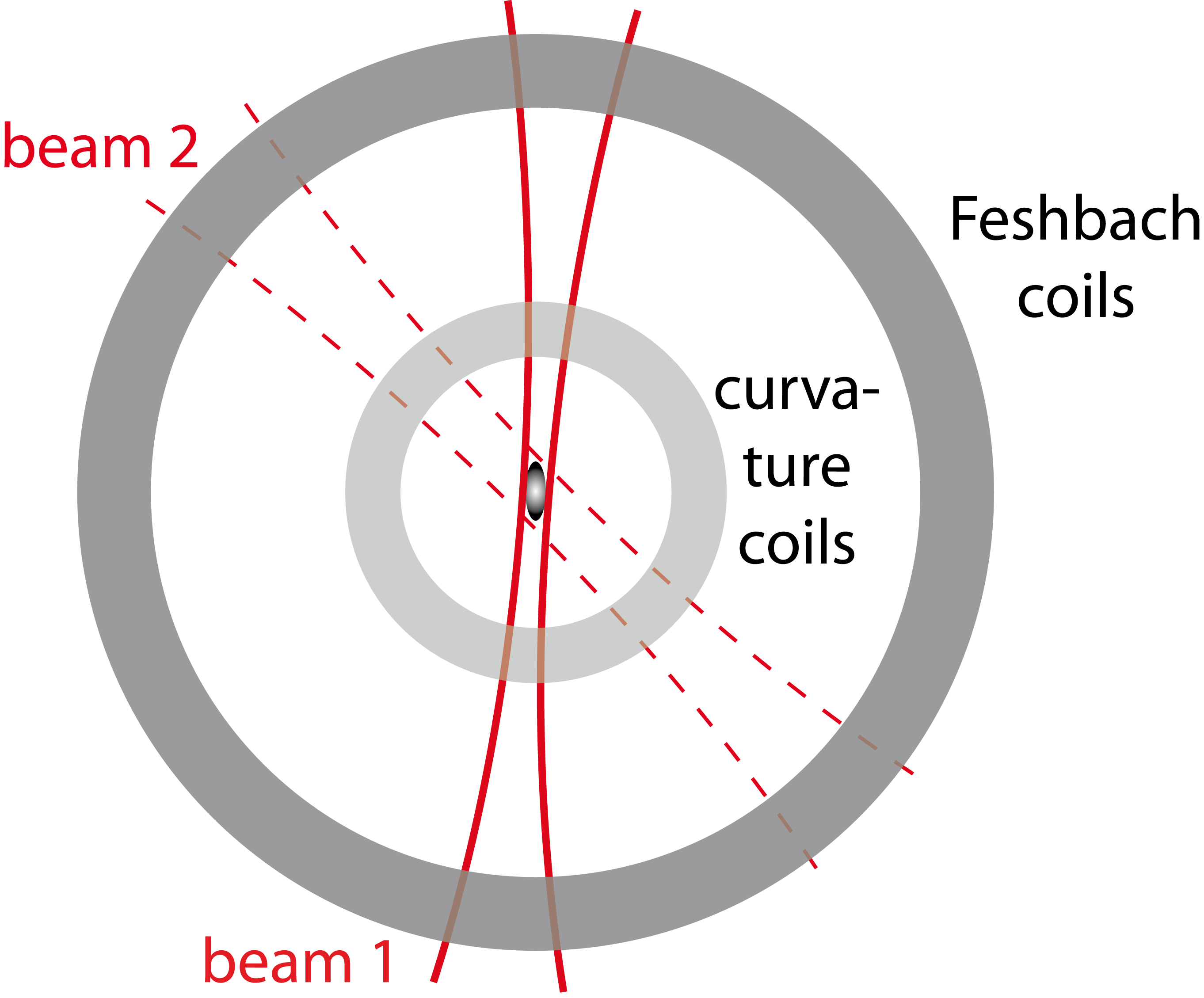}
\caption{Schematic view of the optical trapping beam (beam~1) and the coils for the bias field and magnetic field curvature (shaded areas). Optionally, a second beam (beam~2) can be used for additional axial confinement.}
\label{fig:Traps}
\end{center}
\end{figure}

For further storage and cooling of the atomic cloud, we use a trapping scheme that employs the optical dipole force of an intense infrared laser beam \cite{Grimm2000odt} and the magnetic force in the curvature of the magnetic field \cite{Jochim2003bec}. The latter becomes important when the optical trap is operated with low laser power. A schematic drawing of this hybrid trapping scheme is shown in Fig.~\ref{fig:Traps}.

The principal ODT is formed by a single beam (beam~1) delivered by a 200-W fiber laser that emits light at a central wavelength of 1070\,nm (IPG YLR-200SM-LP). The beam is focused down to a waist of 38\,$\mu$m at the center of the MOT. During loading, the trap is operated with an optical power of 96\,W, which results in a depth of 2.6\,mK for Li. For K the trap depth is larger by a factor of 2.1 and thus amounts to about 5.5\,mK.

Two sets of magnetic field coils are used in our setup to control the bias and curvature field independently; the coils setup is described in more detail in Appendix~\ref{apx:magnets}. For small bias fields the magnetic confinement is very small in the axial direction; here additional axial confinement can be obtained from another infrared beam (beam~2) delivered by a 5-W fiber laser with a central wavelength of 1065\,nm (IPG YLP-5-LP). The single beam is focused to a waist of 97\,$\mu$m and intersects beam~1 at an angle of $53^{\circ}$.

\subsection{Dipole trap loading}
\label{sec:transfer}

\begin{figure}[t]
\begin{center}
\includegraphics[width=0.8\columnwidth]{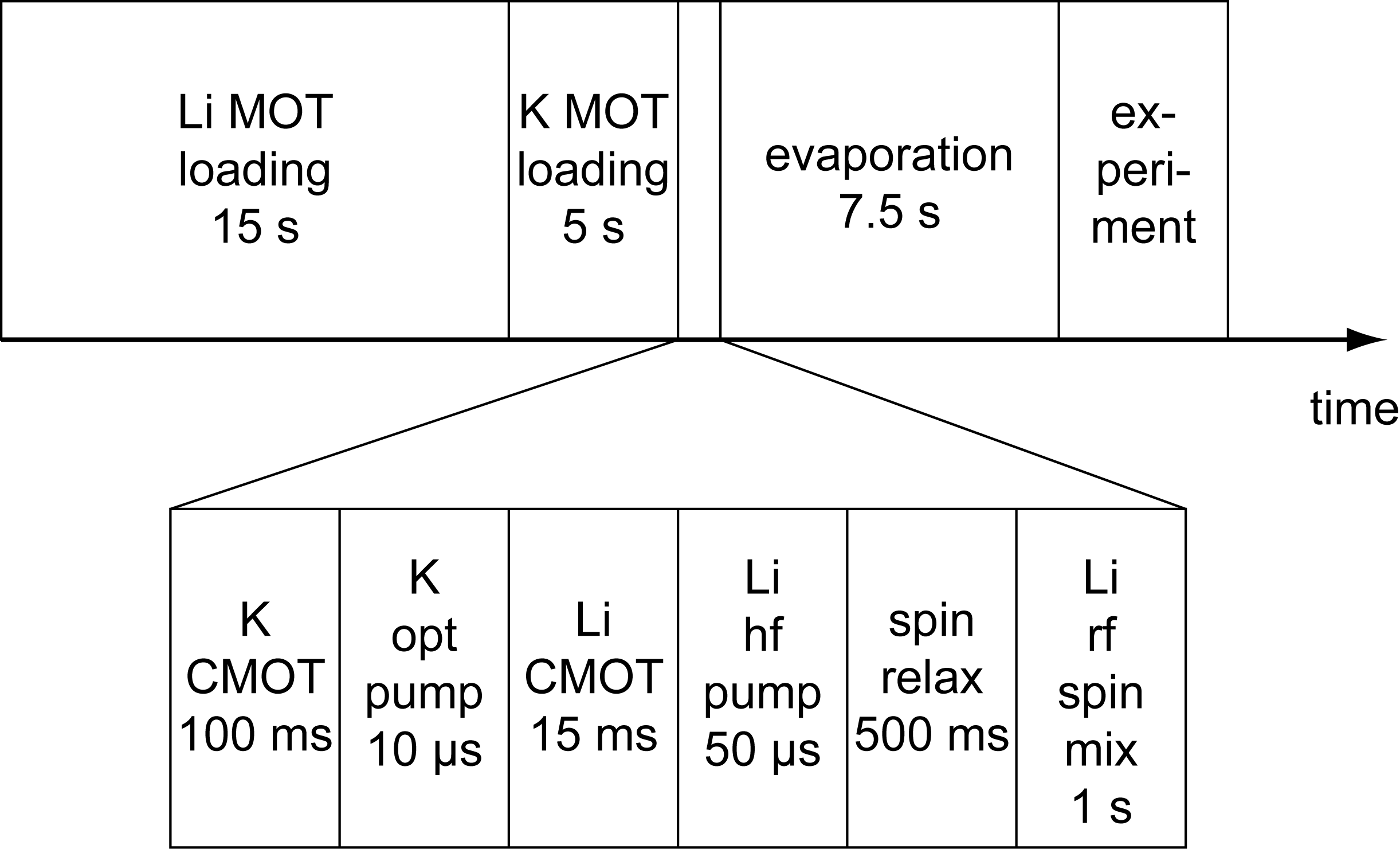}
\caption{Illustration of the loading, transfer and measurement timing sequence. CMOT stands for compressed MOT.}
\label{fig:Sequence}
\end{center}
\end{figure}

Loading cold atoms of a single species from a MOT into a dipole trap is a standard procedure in many experiments. Sub-Doppler cooling and MOT compression are common methods to enhance transfer into the optical trap. The optimized loading of two species, however, needs special procedures. Here we describe the sequential loading scheme that gives us excellent starting conditions for the evaporation of Li and K in a common optical trap. A schematic illustration of the loading and transfer sequence is shown in Fig.~\ref{fig:Sequence}. We found that an optimum is achieved by first transferring K into the optical trap while keeping Li stored in the large-volume, low-density MOT and then performing the Li transfer.

After switching on the ODT, in a first step the K MOT is compressed by ramping up the magnetic field gradient within 50\,ms to 96\,G/cm and bringing the frequencies of the K lasers closer to resonance to a detuning of a few MHz with an intensity lowered to 70\%. At the same time the detuning of the Li MOT is increased to $-47$\,MHz to avoid compression of the Li MOT. At this point the K MOT light is extinguished and the K atoms are confined in the dipole trap while Li is stored practically undisturbed in a MOT at a reduced magnetic field gradient of 64\,G/cm. With the K MOT beams off, untrapped atoms are allowed to escape for 50\,ms.

Potassium has ten Zeeman sublevels in the lowest hyperfine ground state; see Fig.~\ref{fig:LevelScheme}. In order to produce a spin-polarized sample of K in its lowest internal state, we apply an optical pumping scheme that not only transfers the atoms to the lower hyperfine state, but also pumps the atoms to the lowest $m_F$ state.

For optical pumping the quadrupole field is switched off for 2\,ms and only a small guiding field is kept on. Parallel to the field we shine in a $\sigma^-$-polarized laser beam for 10\,$\mu$s, which optically pumps the K atoms into state $|1\rangle$. The optical pumping beam contains two frequency components, one for Zeeman pumping tuned to the ($F=9/2\rightarrow F'=9/2$) transition and another one for hyperfine pumping tuned to the ($F=7/2\rightarrow F'=9/2$) transition as shown in Fig.~\ref{fig:OptPump}. Each frequency component has about 50 times the saturation intensity. During the optical pumping stage, the cloud of Li atoms remains trapped in an optical molasses and can be recaptured without significant losses.

The high-power trapping laser induces a large light-shift on the optical cooling and pumping transitions. Potassium has two optical transitions between the excited 4$^2$P$_{3/2}$ state and the 5$^2$S$_{1/2}$ and 3$^2$D$_{5/2}$ states with wavelengths of 1252\,nm and 1177\,nm, respectively. At high intensities of the ODT these two transitions shift the 4$^2$P$_{3/2}$ level by several 100\,MHz. Therefore optical pumping cannot be performed in the trap and we have to switch it off for a short time. After the 10\,$\mu$s off-time of the ODT needed for the optical pumping, essentially all K atoms are recaptured into the ODT.

At this point we have a sample of a few 10$^4$ polarized \kalium\ atoms in the ODT, surrounded by a magneto-optically trapped cloud of \lithium\ atoms. We finally apply a compressed MOT stage for Li in order to efficiently load this species into the dipole trap. For this, the quadrupole field is ramped back up to 64\,G/cm and the MOT lasers are operated at a very small detuning of $-3$\,MHz from resonance while the power is lowered to 180\,$\mu$W per beam for a duration of 15\,ms. Hyperfine pumping of Li to the lower state is performed by switching the repumping laser off during the last 50\,$\mu$s of the pulse. With this sequence we obtain a few 10$^5$ Li atoms in the lowest two spin states in the ODT at a temperature of about 300\,$\mu$K.

\section{Spin relaxation}
\label{sec:SpinRelax}

A Li$|i\rangle$K$|j\rangle$ mixture can undergo rapid decay via spin relaxation if exoergic two-body collisions can take place that preserve the total projection quantum number $m_{\rm tot}=m_{\rm Li}+ m_{\rm K}=-i+j-4$. In such a process,
\[
\mathrm{Li}|i\rangle+\mathrm{K}|j\rangle\rightarrow\mathrm{Li}|i-1\rangle+\mathrm{K}|j-1\rangle+E_r,
\]
the energy $E_r$ is released. Whenever one of the species is in the absolute ground state, and the other one is in a low-lying state ($i=1$ and $j\leq10$ or $j=1$ and $i\leq3$), spin relaxation is strongly suppressed \cite{Simoni2003mco}.

Since optical Zeeman pumping does not lead to a perfect transfer of all K atoms into the lowest spin state, we exploit spin relaxation to fully spin polarize K into state \1. We investigated the conditions under which spin relaxation can be used for this purpose. In these measurements we apply only hyperfine pumping, but no Zeeman pumping. We start with an almost equal mixture of Li in its lowest two hyperfine states, \1\ and \2, trapped together with a population of K in all Zeeman substates $j\leqslant10$.

\begin{figure}[tb]
\begin{center}
\includegraphics[width=\columnwidth]{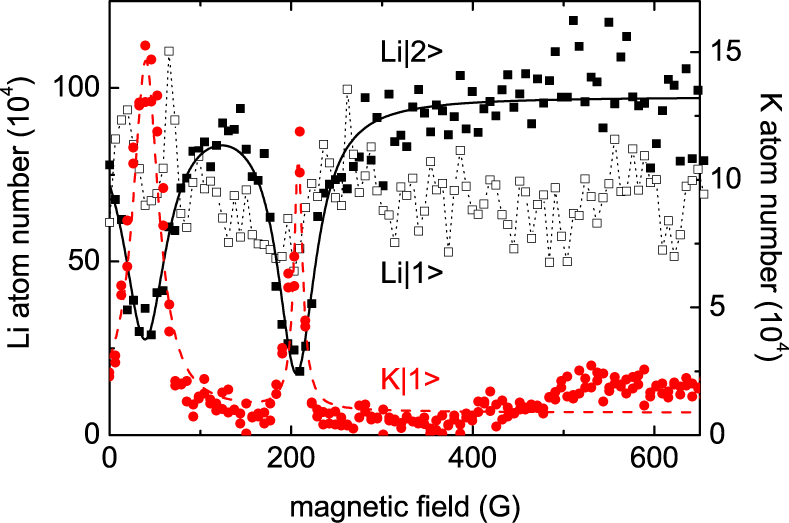}
\caption{Magnetic field dependence of spin relaxation. The numbers of atoms in different spin states are measured by state-selective absorption imaging after a storage time of 500\,ms in the dipole trap. The filled (open) squares give the Li\2\ (\1) atom number, the filled circles are K\1. The two pronounced features that are visible at 40\,G and 207\,G are fitted by Lorentzians to determine their positions and widths.}
\label{fig:SpinRelaxBfield}
\end{center}
\end{figure}

We investigate the magnetic field dependence of the spin relaxation by holding the sample for 500\,ms at a variable magnetic field. The trap is operated under the same conditions as during trap loading, i.e. with a trap depth of 2.6\,mK for Li and 5.5\,mK for K. The atom numbers are measured using spin-selective absorption images, which are always taken at a bias field of 1190\,G \cite{endnote:HighFieldImaging}. Figure~\ref{fig:SpinRelaxBfield} shows the resulting atom numbers of Li in states \1\ (open squares) and \2\ (filled squares) as well as K in state \1\ (filled circles). Two distinct peaks in the K\1\ atom number are visible at 40\,G and 207\,G and coincide with dips in the Li\2\ atom number. These features are fitted with Lorentzians to determine their positions and widths.

The release energy $E_r$ at 40\,G (207\,G) corresponds to 2.1\,mK (5.8\,mK). For an inelastic collision between two atoms with different masses, the resulting kinetic energy contributions are inversely proportional to the mass. For the \lithium-\kalium\ combination, 87\% of the released energy is transferred to the lighter Li atoms (mass $M_\mathrm{Li}$) and only 13\% to the heavier K (mass $M_\mathrm{K}$). A necessary condition for the trap depth $U_\mathrm{K}$ under which a K atom stays confined is
\begin{equation}
U_\mathrm{K} > \frac{M_\mathrm{Li}}{M_\mathrm{K} + M_\mathrm{Li}} E_r,
\end{equation}
and analogously for $U_\mathrm{Li}$. The mass factor along with the about two times larger trap depth for K explains why we observe loss of Li atoms during the spin relaxation while K stays trapped. Furthermore, a K atom in a Zeeman level higher than \2\ will need multiple collisions with Li\2\ in order to fully polarize. That explains why much more Li\2\ is lost than K\1\ is gained during this process.

\begin{figure}[tb]
\begin{center}
\includegraphics[width=0.975\columnwidth]{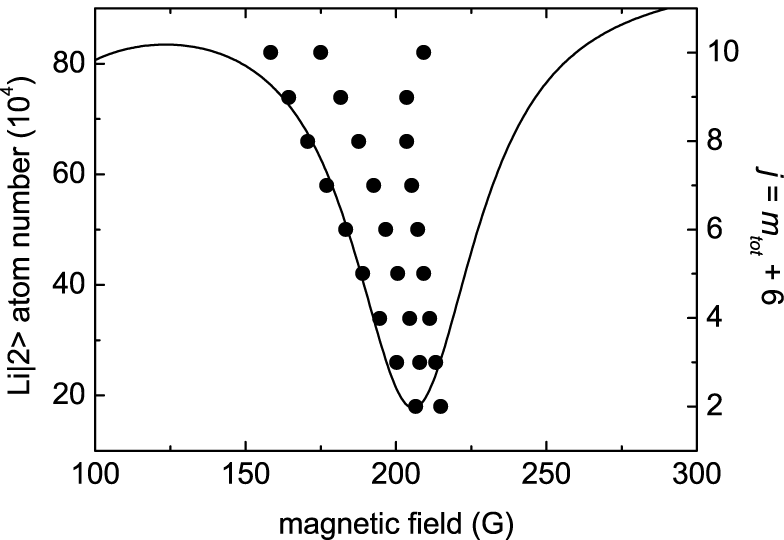}
\caption{Interpretation of the 207\,G spin relaxation feature in terms of Feshbach resonances. The dots show the calculated positions of \fbr s between Li\2\ and K$|2\leq j\leq10\rangle$ \cite{Tiecke2009fri}. Also plotted is the Lorentzian fit to the Li\2\ loss feature from Fig.~\ref{fig:SpinRelaxBfield} for comparison.}
\label{fig:Li2KsWave}
\end{center}
\end{figure}

We interpret our data by comparing the position of the two spin relaxation features with the location of known interspecies \fbr s, since we expect enhanced inelastic loss close to a \fbr\ \cite{Chin2009FBR}. As Fig.~\ref{fig:Li2KsWave} shows, there is a series of $s$-wave \fbr s between Li\2\ and K$|2\leq j\leq10\rangle$ near the 207\,G feature. The distribution of \fbr s coincides with the width of the observed spin relaxation feature. Note that the experiment is performed at relatively high temperature causing considerable broadening of the \fbr s. Therefore individual resonances cannot be resolved.

For the feature of enhanced spin relaxation at 40\,G, there are no interspecies $s$- or $p$-wave \fbr s and thus we cannot explain it by means of scattering resonances. However, at low magnetic fields $B$ the release energy $E_r$ increases $\propto B$, which leads to a corresponding increase in the density of continuum states in the decay channel. We speculate that the corresponding threshold behavior may explain the increase at lower fields. Then, already at a few ten Gauss the nuclear spin of Li decouples from the electron spin, which may lead to a reduction of loss.

In a second set of experiments we investigate the time scale on which spin relaxation occurs at the two relevant magnetic fields 40\,G and 207\,G. For both fields we find that the time scale for the process is 150\,ms and a steady state is essentially reached after 500\,ms.

Spin relaxation is a very efficient process that allows us to fully polarize our K sample without loss of K atoms. Since initially much more Li atoms are present in the trap, the Li loss is a minor problem. The resulting imbalance of the two Li spin states can be removed by driving radio-frequency (rf) transitions between the two states.

\section{Evaporation and sympathetic cooling}
\label{sec:Evaporation}

A spin-mixture of Li\1\ and Li\2\ near the broad 834-G \fbr\ facilitates highly efficient evaporative cooling, as it is well known in the field of strongly interacting Fermi gases \cite{OHara2002ooa, Inguscio2006ufg}. The efficiency of the cooling process is due to the very favorable combination of a large elastic scattering cross section with very low losses. In Ref.~\cite{Spiegelhalder2009cso} we have already demonstrated the possibility of using the \lithium\ spin-mixture as an efficient cooling agent to sympathetically cool another species. In this way we have demonstrated the attainment of a double-degenerate Fermi-Fermi mixture of \lithium\ and \kalium. Here we present additional information on the experimental procedures, and the combined evaporative and sympathetic cooling process.

Let us first summarize our main findings of Ref.~\cite{Spiegelhalder2009cso} on the collisional stability of the three-component Fermi gas of Li in the lowest two spin states together with K in the lowest spin state. Interspecies collisional loss is negligible on the BCS side of the broad Li resonance ($B>900$\,G) and quite weak even exactly on resonance (834\,G). Substantial loss, however, occurs on the BEC side of the resonance in a range between about 650 and 800\,G. The latter is a result of inelastic collisions between K atoms with weakly bound Li dimers. Consequently, the combined evaporation and sympathetic cooling process needs to be performed on the BCS side of the Li resonance. For our experiments we choose a field of 1190\,G. Here the Li scattering length is $-2900~a_0$ and the interspecies scattering length is about $+60~a_0$.

\begin{figure}[t]
\begin{center}
\includegraphics[width=\columnwidth]{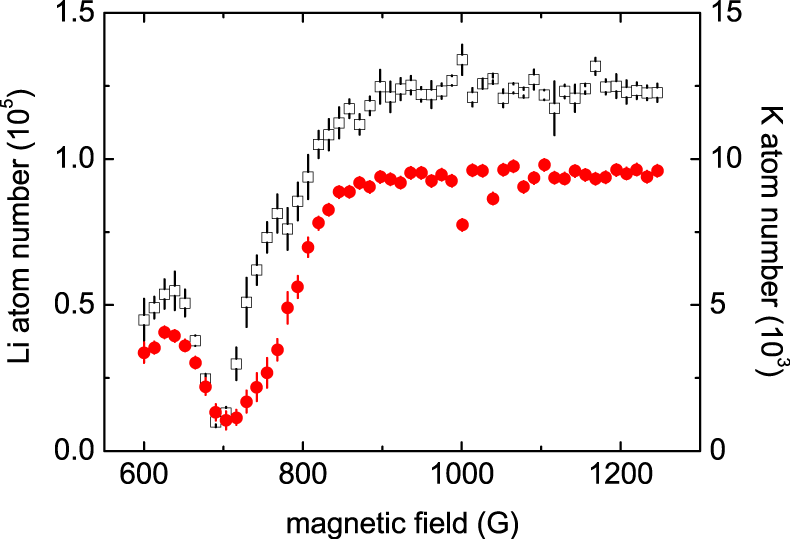}
\caption{Li and K atom numbers after evaporation performed at variable magnetic field. Open squares show the number of Li atoms per state, filled circles show the K atom number.}
\label{fig:EvapBField}
\end{center}
\end{figure}

Before starting the evaporation process, we carefully balance the population of the two spin states Li\1\ and Li\2. This is particularly important in cases when the spin relaxation stage has caused considerable losses in \2. The spin balance is accomplished by driving the rf transition $|1\rangle\leftrightarrow|2\rangle$ using a series of 20 ramps over 10\,kHz with a duration of 50\,ms each \cite{Strecker2003coa}. This procedure is performed at 1190\,G, where the Li spin mixture is outside of the strongly interacting regime and interaction-induced rf line shifts are relatively small. Note that the evaporation process is much more sensitive to a spin imbalance on the BCS side of the resonance than on the BEC side of the resonance. The reason is that in the latter case the molecule formation can lead to a self-balancing of the spin population during evaporation \cite{Grimm2008Varenna}.

Evaporation of the mixture is performed in the principal ODT, beam 1. The evaporation ramp consists of two stages, technically implemented in different ways. In the first stage, we use a digital input of the laser control unit to reduce the ODT power to about 15\,W. This ramp is linear and takes 1.5\,s. In a second stage, an acousto-optical modulator (AOM) is used to decrease the power in a nearly exponential ramp. The evaporation ramp is typically stopped after 6\,s when the laser power is 60\,mW. At this point the trap frequencies  in the radial directions are 394\,Hz for Li and 219\,Hz for K. In the axial direction the trap frequency is dominated by the magnetic confinement and is 27\,Hz for Li and 11\,Hz for K.

The experimental data in Fig.~\ref{fig:EvapBField} show the number of atoms remaining after the complete evaporation ramp for a magnetic field varied across the full tuning range offered by the Li \fbr. The data correspond to the observations of Ref.~\cite{Spiegelhalder2009cso}, showing pronounced loss on the BEC side of the Li \fbr\ and large stability on its BCS side. In the high-field region between 950\,G and 1250\,G, where the Li scattering length varies between $-5300~a_0$ and $-2800~a_0$, the magnetic field has no significant influence.

\begin{figure}[t]
\begin{center}
\includegraphics[width=0.975\columnwidth]{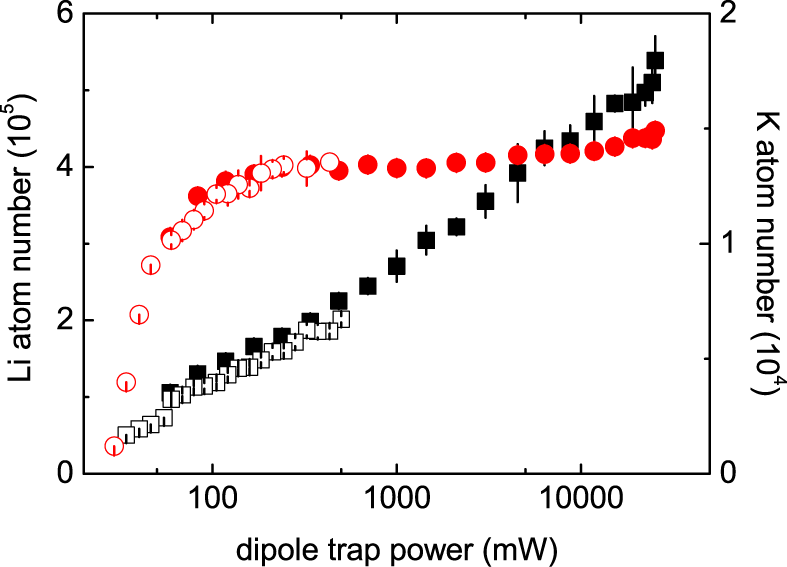}
\caption{Evolution of the atom numbers during the second stage of evaporation. The Li atom number per state is plotted using squares while circles represent the K atom number. Filled (open) symbols represent data from measurements using fluorescence (absorption) imaging.}
\label{fig:EvaporationPower}
\end{center}
\end{figure}

In order to analyze the cooling process, we stop the evaporation ramp at a variable endpoint and measure the number of Li and K atoms. The measurements are performed by recapture into the MOT and subsequent detection of the fluorescence intensity or, at lower power, by absorption imaging after release from the ODT into free space. Figure~\ref{fig:EvaporationPower} shows that the Li atom number steadily decreases while the K cooling proceeds essentially without losses. Note that the trap depth for K is a factor of 2.1 larger than for Li. This changes at about 100\,mW, as the gravitational sag of K reduces the trap depth and we begin to see significant loss of K when further lowering the power of the ODT.

\begin{figure}[tb]
\begin{center}
\includegraphics[width=\columnwidth]{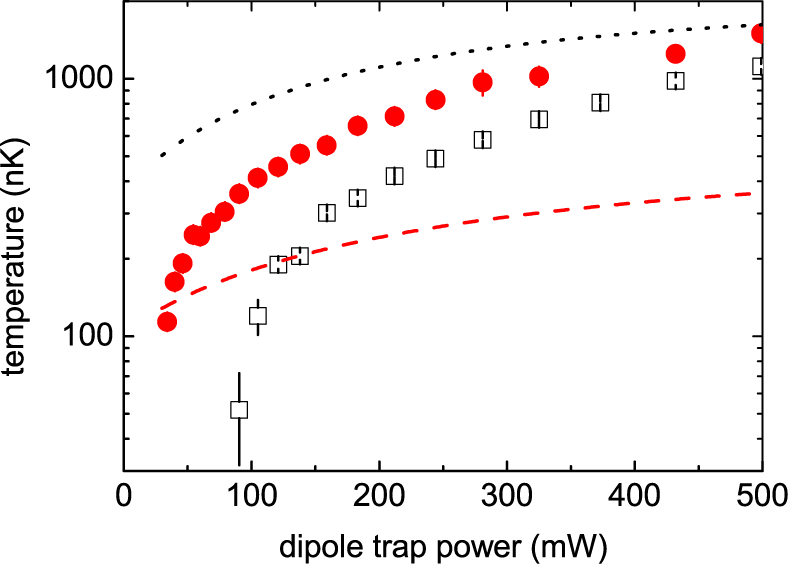}
\caption{Temperature evolution during the last part of the evaporation. Open squares (filled circles) indicate the Li (K) temperature. Also plotted are curves that represent the evolution of the Fermi temperature for Li (dotted line) and K (dashed line).}
\label{fig:Temp}
\end{center}
\end{figure}

Figure~\ref{fig:Temp} shows the temperature evolution of Li and K in the last part of the evaporation ramp. We extract the Li temperature by fitting a Thomas-Fermi profile to absorption images of the atomic cloud after \tof. The K temperature is determined using a simple Gaussian fit, as the sample here remains in the non-degenerate regime. Throughout the whole evaporation the temperature of K lags behind the temperature of Li.

At the end of an extended evaporation ramp, at a trap power of 40\,mW, the radial trap frequencies for Li (K) amounts to 320\,Hz (160\,Hz). In the axial direction the trap frequency is dominated by the magnetic confinement and is 27\,Hz for Li and 11\,Hz for K. The Fermi temperatures are calculated to be $T^\mathrm{Li}_F=390$\,nK for Li and $T^\mathrm{K}_F=135$\,nK for K. Following the scheme we have presented in Ref.~\cite{Spiegelhalder2009cso}, we continue cooling of the mixture by holding it in this shallow trap for 5\,s. This way we achieve a final K temperature of 50\,nK corresponding to a degeneracy of $T^\mathrm{K}/T_F^\mathrm{K}\approx 0.6$ while Li is deeply degenerate with $T^\mathrm{Li}/T_F^\mathrm{Li}<0.2$.

Since Li is the coolant of our evaporation scheme, we adjust the amount of K with which evaporation starts such that at the end of evaporation we have about ten times more atoms of Li in each spin state than atoms in K\1. In this situation K can be used as a probe for the Li mixture. One example of this idea has already been presented in Ref.~\cite{Spiegelhalder2009cso}. The measurement of the K temperature was used in order to get a firm upper bound for the temperature of the Li bath even in the strongly interacting regime. This method was recently adopted to using $^7$Li as a probe in Ref.~\cite{Nascimbene2009ett}.

\section{Preparation near interspecies \fbr s}
\label{sec:StatePrep}

The \lithium-\kalium\ mixture offers several $s$-wave \fbr s in the range between 150\,G and 200\,G \cite{Wille2008eau, Tiecke2009sfb}. All of them tend to be quite narrow, which is a common situation in cases of moderate values of the background scattering length \cite{Chin2009FBR}. The broadest resonances were found for the channels Li\1K$|7...10\rangle$ with widths between 1\,G and 2\,G \cite{Tiecke2009fri}. The energetically lowest channel Li\1K\1, which is of particular interest because of the energetic suppression of any two-body decay, features two resonances with calculated widths around 100\,mG. In this work, we focus on the resonance near 168\,G. We show how a degenerate two-component Li-K mixture can be prepared near this resonance after sympathetic cooling at high magnetic field and present measurements on inelastic and elastic properties of the mixture.

\begin{figure}[t]
\begin{center}
\includegraphics[width=\columnwidth]{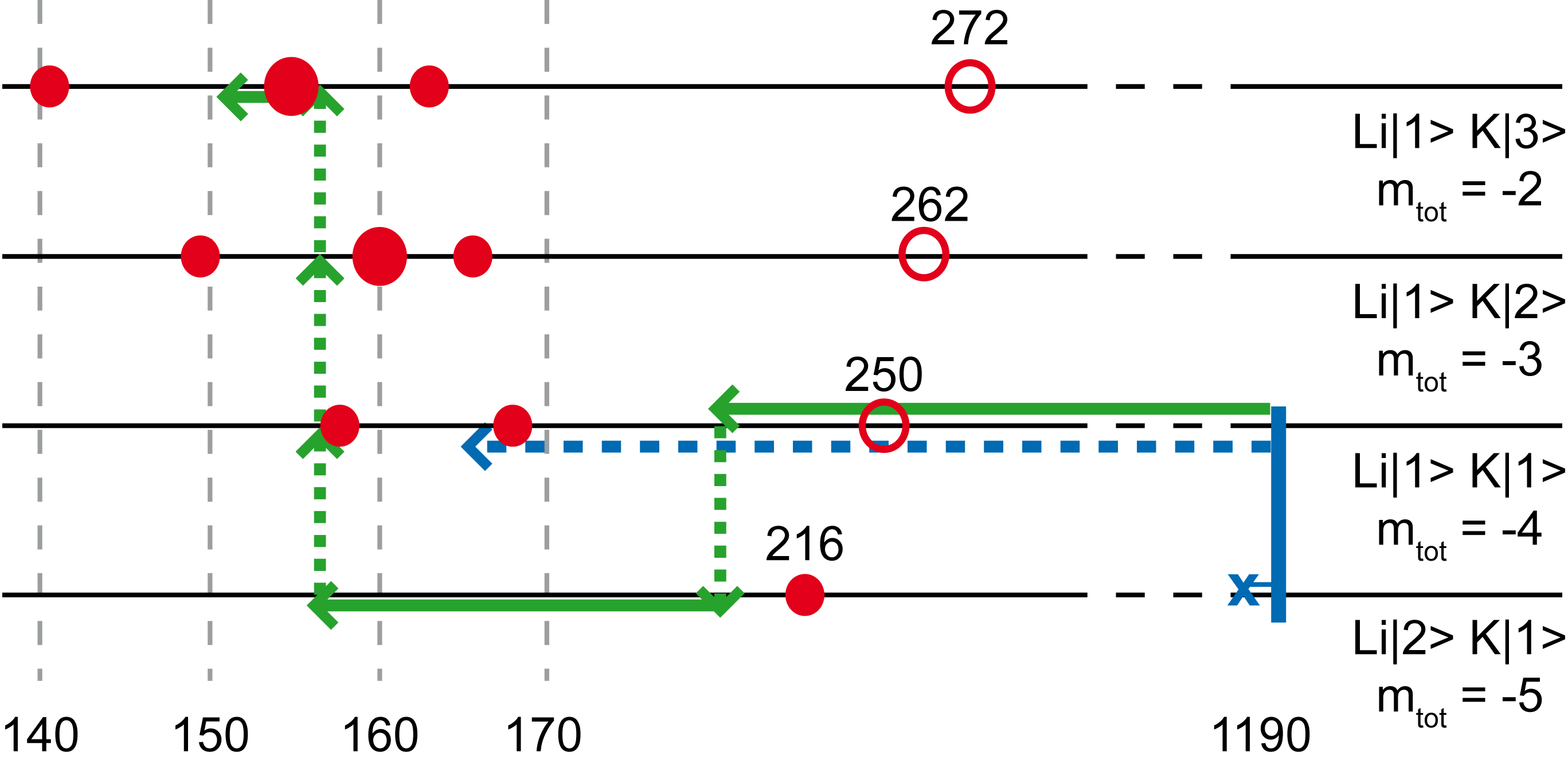}
\caption{Preparation scheme of a two-component mixture near interspecies \fbr s. The horizontal lines indicate four different Li-K spin channels being of particular relevance for our experiments. The numbers give the magnetic field values in Gauss. Filled (open) circles represent $s$-wave ($p$-wave) interspecies resonances. The Li\1\ and Li\2\ intraspecies $p$-wave resonances (not shown) are located at 160\,G and 215\,G respectively \cite{Zhang2004pwf, Schunck2005fri} and nearly coincide with interspecies resonances. For the $s$-wave resonances, the relative widths are indicated by the size of the symbols. State transfer, indicated by vertical dashed lines, is achieved by rf transitions. After the evaporation at 1190\,G, the Li\2\ population is removed by a resonant laser pulse, as indicated by the \textbf{x}.}
\label{fig:Ramp2FBR}
\end{center}
\end{figure}

When ramping down the magnetic field from its evaporation setting (1190\,G) to the interspecies resonances, one has to cross the region of the broad 834-G Li\1\2\ \fbr. If this spin channel is populated, the formation of $^6$Li dimers inevitably leads to strong losses from the atomic sample. To avoid this problem, we remove one of the Li spin components by the light pressure of a resonant laser pulse \cite{Du2008ooa} before starting the ramp. Note that already the momentum kick from one photon is sufficient to push a Li atom out of the shallow trap after evaporation. The pulse is applied for 10\,$\mu$s with a few times the saturation intensity. We find that at 1190\,G the interaction between the two spin components is weak enough to avoid any significant effect on the population in the remaining spin state. In this way, we reduce the three-component Fermi-Fermi gas to a two-component mixture.

To approach a specific interspecies resonance, it is also important to avoid the effect of other inter- and intraspecies resonances. We find that our ramps are fast enough (ramp speed up to 20\,G/ms) to cross all the $p$-wave resonances without any problem. However, we find substantial losses on the interspecies $s$-wave resonances, even on the weaker ones. This already points to efficient molecule association \cite{Chin2009FBR} as we will discuss in Sec.~\ref{sec:Molecules}.

Figure~\ref{fig:Ramp2FBR} illustrates the procedures applied to reach specific interspecies Feshbach resonances. While it is straightforward to reach the 168-G resonance in the Li\1K\1\ channel by a fast ramp after removal of the state Li\2, other resonances require more elaborate methods. As an example, we discuss the 155-G resonance in the Li\1K\3\ channel, which is of interest as one of the broadest resonances (width between 0.5 and 1\,G) in the low-lying spin channels. Here a possible way is to transfer the Li atoms from \1 to \2 after ramping down the field to $\sim$200\,G, thus converting the sample into a Li\2K\1 mixture. This can be done with very high efficiency using rf transfer methods. Then the ramp is continued down to a value close to 155\,G and three subsequent rf transfers are applied to convert the population from Li\2K\1\ to Li\1K\3. This procedure avoids all detrimental resonances. Analogous schemes can be applied to reach any other desired resonance.

\begin{figure}[t]
\begin{center}
\includegraphics[width=0.85\columnwidth]{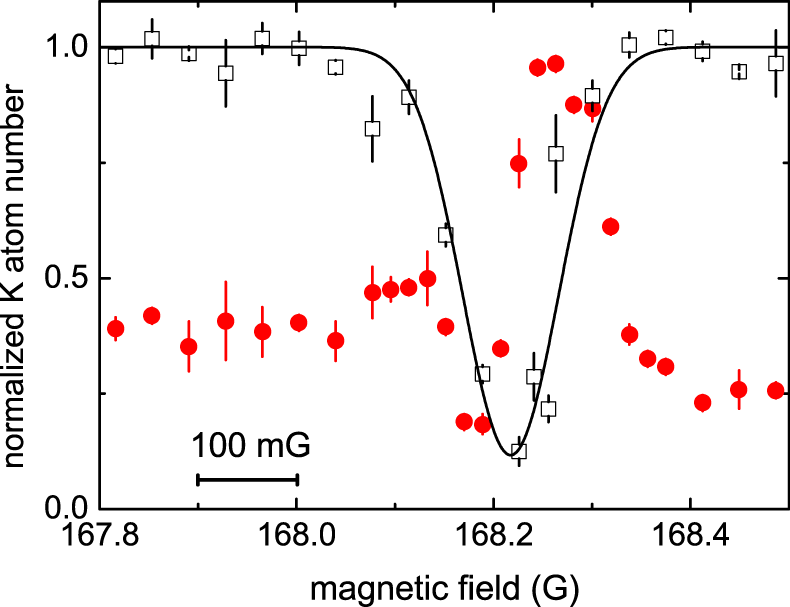}
\caption{Loss measurement and evaporative cooling near an interspecies \fbr. Plotted is the K atom number normalized to the background value of the loss measurement away from resonance. The open squares show a set of loss measurements holding the sample at variable magnetic field for 5\,s. The solid line is a Gaussian fit to the data. The filled circles show a corresponding set of measurements where evaporation was performed by lowering the optical power to one third of its initial value within 3\,s.}
\label{fig:LiKFBRevap}
\end{center}
\end{figure}

In a set of experiments performed at the 168-G interspecies \fbr\ in the Li\1K\1\ channel, we investigate aspects of inelastic and elastic collisions. Initially, we prepare about $2\times10^5$ Li atoms together with about $1.4\times10^4$ K atoms at a temperature of about 300\,nK. The power of the trapping beam (beam 1) is 170\,mW, corresponding to a radial (axial) trap frequency of 660\,Hz (14\,Hz) for Li. For K the trap frequencies are 375\,Hz and 6\,Hz respectively. The peak densities of the clouds are $n^{\rm Li}_0\approx 2\times 10^{12}\,$cm$^{-3}$ and $n^{\rm K}_0\approx 4\times 10^{11}\,$cm$^{-3}$ and the degeneracies are $T^{\rm Li}/T_F^{\rm Li}\approx0.3$ and $T^{\rm K}/T_F^{\rm K}\approx1.5$. Note that these conditions are deliberately prepared with an incomplete evaporation ramp, stopped at 170\,mW instead of the usual final power of 60\,mW (Sec.~\ref{sec:Evaporation}).

In the first series of measurements, we ramp the magnetic field to a variable value and study the loss of atoms after a hold time of 5\,s. For detection, the remaining atoms are recaptured into the two-species MOT and their fluorescence is recorded. The K atom number, normalized to the background value away from resonance, is plotted in Fig.~\ref{fig:LiKFBRevap} (open squares). We observe a loss feature centered at 168.22\,G. Ramping across the \fbr\ does not lead to loss, indicating that the phase-space density used in these experiments is insufficient for adiabatic molecule creation during the magnetic field ramp.

In the second series of measurements, we investigate whether an effect of enhanced elastic collisions can be observed in evaporative cooling near the interspecies resonance. Here we lower the power of beam 1 to 55\,mW within 3\,s, which results in a radial (axial) trap frequency of 375\,Hz (14\,Hz) for Li and 210\,Hz (5\,Hz) for K. As before, the number of remaining atoms is determined by recapture into a MOT and fluorescence detection. The corresponding data, plotted in Fig.~\ref{fig:LiKFBRevap} (filled circles), show a pronounced asymmetry and thus a different behavior on the two sides of the Feshbach resonance. On its high-field side, corresponding to large negative scattering length, we observe a maximum in the recaptured atom number at 168.26\,G. This signifies evaporative cooling with an enhanced elastic scattering cross section as compared to the background value. At lower fields, however, loss enhanced by the resonance dominates and leads to a minimum in atom number at 168.18\,G. The loss properties on the two sides of the Feshbach resonance are thus found to be strikingly different with more favorable conditions on the side of negative scattering length, where no weakly bound molecular state exists.

\begin{figure}[t]
\begin{center}
\includegraphics[width=0.85\columnwidth]{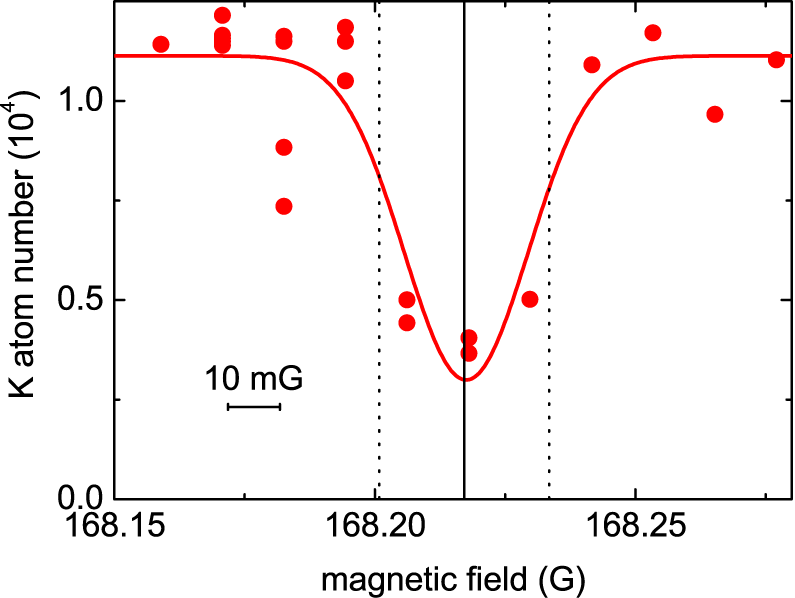}
\caption{Scan of the Li-K \fbr\ at 168\,G with 10\,ms hold time. The $1/e$-width of the loss feature (indicated by the dotted lines) is determined by fitting a Gaussian to the experimental data and amounts to 33\,mG.}
\label{fig:FastLoss}
\end{center}
\end{figure}

To determine the resonance position more precisely, we prepare a sample with higher phase-space density than used for the previous sets of experiments. Now both beams of the ODT are used. Beam 2 is held at a constant power of 250\,mW, corresponding to a trap depth of 1\,$\mu$K for Li and 2.1\,$\mu$K for K. Its purpose is to add confinement along the weak direction of beam 1 at the very end of evaporation. Evaporation at 1190\,G and the ramp to low magnetic field proceed as described above. For further cooling, we create a balanced mixture of Li\1\ and \2\ at 170.5\,G, using a sequence of rf sweeps. Afterwards, beam 1 is ramped from a Li trap depth of 1.9\,$\mu$K to 1.5\,$\mu$K during one second and the sample is left to thermalize for another second. Then Li\2\ is removed by a short pulse of resonant light. At this point the sample has a temperature of $\sim$200\,nK and contains about $1.3\times10^4$ K atoms and $8\times10^4$ Li atoms. With the trap oscillation frequencies of axially 90\,Hz (50\,Hz), and radially 390\,Hz (210\,Hz) for Li (K), we calculate Fermi temperatures of about 900\,nK for Li and 270\,nK for K, corresponding to $T^{\rm Li}/T_F^{\rm Li}\approx0.2$ and $T^{\rm K}/T_F^{\rm K}\approx0.7$. The K cloud has less than half the size of the Li cloud. For both components the density in the center of the trap is about $2\times 10^{12}\,$cm$^{-3}$.

Under these deep cooling conditions, we detect the fast atom loss as a function of the magnetic field. In order to approach the magnetic field value of interest without forming molecules, K is transferred into state \2\ by an rf $\pi$-pulse prior to the magnetic field ramp. At the final field, K is transferred back to state \1\ by another $\pi$-pulse. After a hold time of 10\,ms, the remaining K atom number is measured. Figure~\ref{fig:FastLoss} shows the corresponding data. We observe maximum loss of atoms centered at 168.217\,G, with an estimated calibration uncertainty of 10\,mG.

\section{Creation of ultracold Fermi-Fermi molecules}
\label{sec:Molecules}

Here, we describe our basic methodology for molecule creation and detection (Sec.~\ref{sec:CreationAndDetectionSchemes}), present experimental results (Sec.~\ref{sec:ExperimentalResults}) and discuss our findings (Sec.~\ref{sec:Discussion}).

\subsection{Creation and detection schemes}
\label{sec:CreationAndDetectionSchemes}

\begin{figure}[b]
\begin{center}
\includegraphics[width=\columnwidth]{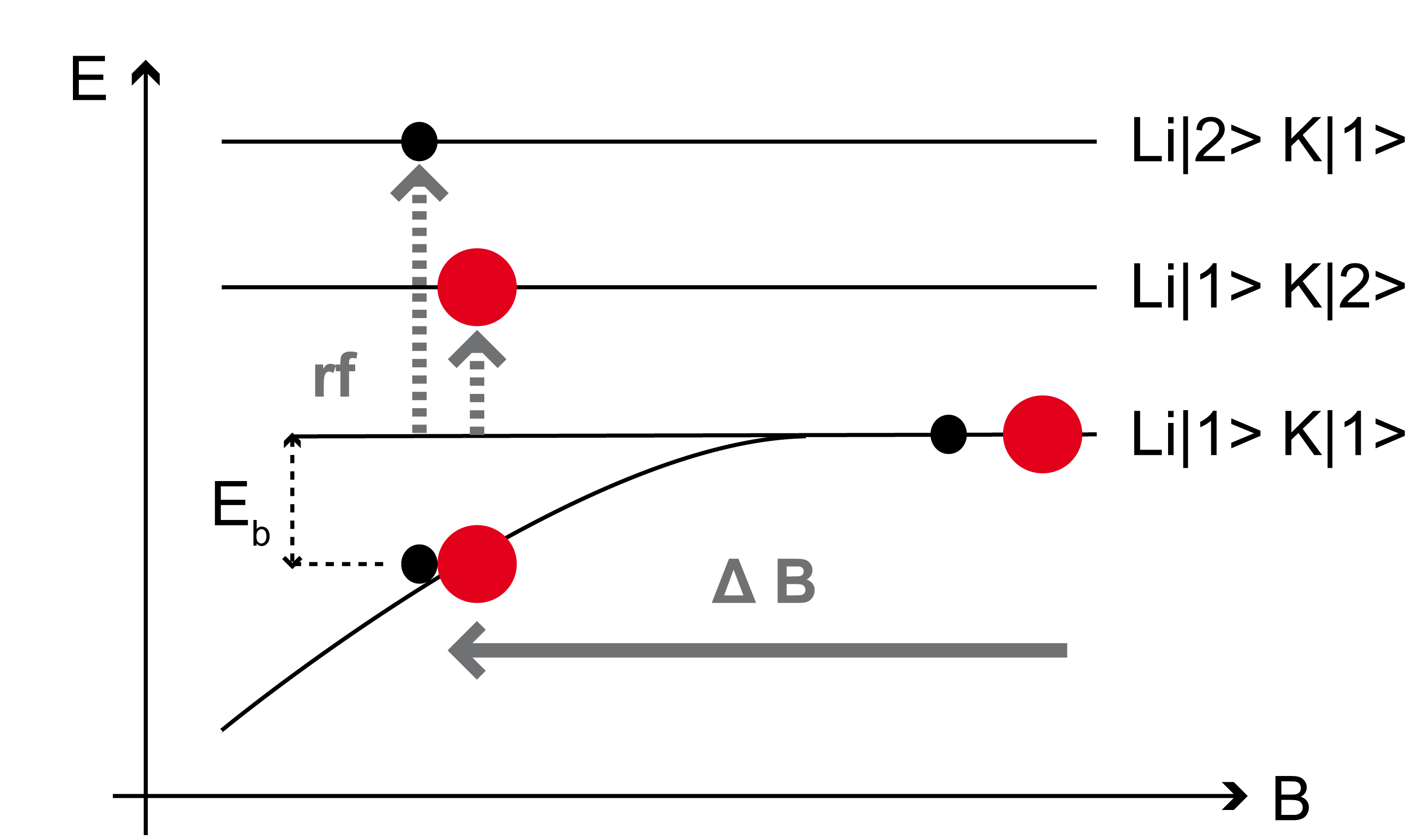}
\caption{Molecules are associated by a magnetic field ramp, indicated by the arrow labeled $\Delta B$, across a \fbr. Transitions to higher atomic spin states are driven by rf pulses. Atoms bound in a molecule are not affected because of the binding energy $E_b$.}
\label{fig:ImageRF}
\end{center}
\end{figure}

The creation of the molecules starts with a Li\1K\1\ mixture under the same conditions as prepared for Fig.~\ref{fig:FastLoss}. The molecules are associated by a magnetic field ramp from 170.5\,G to 168.19\,G within 10\,ms, crossing the Li\1K\1 168-G \fbr\ (ramp speed 0.23\,G/ms). Instantly after the ramp, the sample is released from the ODT.

Selective imaging of molecules and remaining unpaired atoms is possible after transfer of the unpaired atoms to the states Li\2\ and K\2. An rf $\pi$-pulse, tuned to the atomic \1$\rightarrow$\2\ transition, is used for this purpose \cite{endnote:RFTransitions}; see Fig.~\ref{fig:ImageRF}. Atoms bound in LiK molecules are not transferred to state \2\ if the molecular binding energy detunes the transition far enough from the free atom transition to be outside of the Fourier spectrum of the rf pulse. This condition requires a detuning of 23\,kHz for K, which is reached 9\,mG below resonance according to the relative magnetic moment of the molecular state \cite{Tiecke2009fri}. The rf pulses are applied one after the other during the 0.4\,ms free expansion of the sample.

\begin{figure}[t]
\begin{center}
\includegraphics[width=\columnwidth]{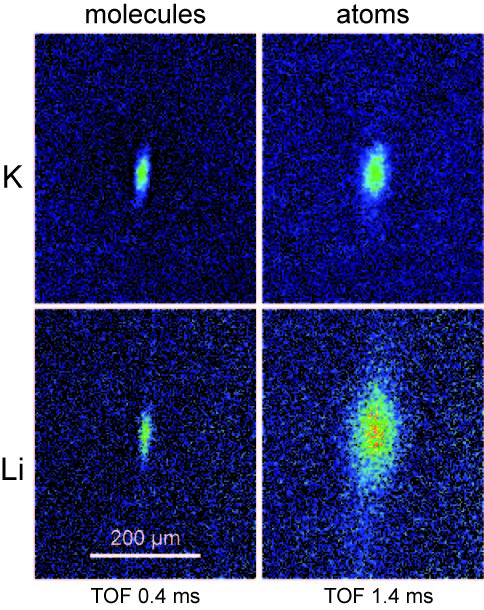}
\caption{Absorption images of LiK Feshbach molecules and unpaired atoms taken at a magnetic field of 168.19\,G. The upper row shows images of molecules and atoms taken with light resonant to the K transition whereas the lower row shows images taken with light resonant to the Li transition. The left column shows molecules imaged after 0.4\,ms \tof\ (TOF) expansion and the right column unpaired atoms imaged 1\,ms later.}
\label{fig:LiKMolecules}
\end{center}
\end{figure}

State-selective absorption images are taken simultaneously on cycling transitions starting from the Li\1\ and K\1\ states. This way, molecules are imaged directly. The resulting pictures are shown in the left-hand column of Fig.~\ref{fig:LiKMolecules} \cite{endnote:MoleculeImaging}. A second pair of images, this time of the unpaired atoms, which have been transferred to the \2\ states, is taken 1\,ms later and shown in the right-hand column of Fig.~\ref{fig:LiKMolecules} \cite{endnote:unpairedAtomImaging}.

Absorption imaging of the molecules gives lower boundaries $\mathcal{N}^{\rm K}_{\rm mol}$ and $\mathcal{N}^{\rm Li}_{\rm mol}$ for the real molecule numbers $N^{\rm K}_{\rm mol}$ and $N^{\rm Li}_{\rm mol}$ since the absorption cross section of atoms bound in LiK molecules is somewhat smaller than the one of unpaired atoms. Close to the \fbr\ the cross section is similar to the one of free atoms and decreases for increasing binding energy. The number of remaining unpaired atoms $N^{\rm K}_{\rm free}$ and $N^{\rm Li}_{\rm free}$ can be obtained from the second pair of absorption images.

From K images (top row of Fig.~\ref{fig:LiKMolecules}) we obtain $\mathcal{N}^{\rm K}_{\rm mol}=3\times10^3$ and $N^{\rm K}_{\rm free}=9\times10^3$ and from Li images (bottom row of Fig.~\ref{fig:LiKMolecules}) $\mathcal{N}^{\rm Li}_{\rm mol}=4\times10^3$ and $N^{\rm Li}_{\rm free}=8\times10^5$. The small cloud of K is immersed in a much larger degenerate Li bath. The molecule conversion efficiency is therefore best characterized by the K conversion efficiency. A lower bound for the molecule fraction can be determined from K absorption images as $\mathcal{F}=\mathcal{N}^{\rm K}_{\rm mol}/(\mathcal{N}^{\rm K}_{\rm mol}+N^{\rm K}_{\rm free}$). From the images shown in Fig.~\ref{fig:LiKMolecules} we obtain $\mathcal{F}=0.25$.

\subsection{Experimental results}
\label{sec:ExperimentalResults}

\begin{figure}[tb]
\begin{center}
\includegraphics[width=0.85\columnwidth]{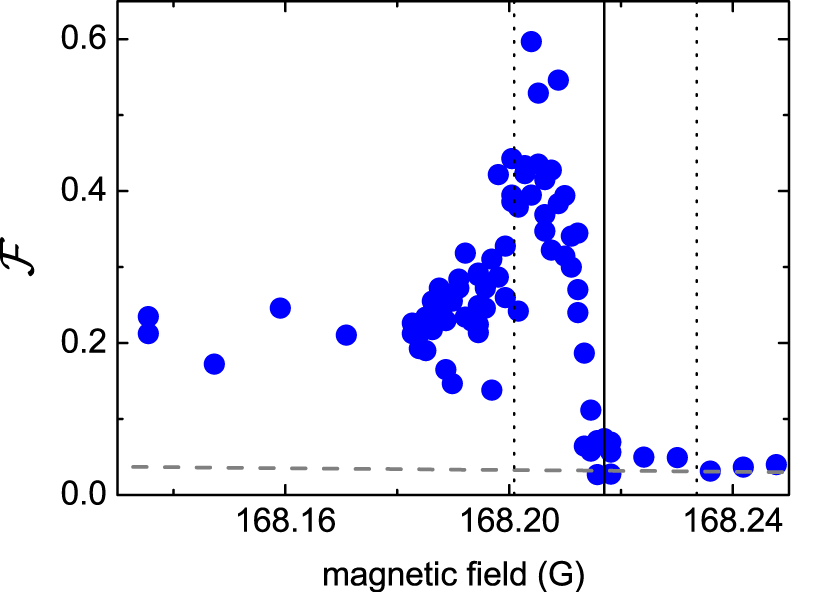}
\caption{Lower bound of molecule fraction $\mathcal{F}$ in dependence of the final magnetic field value of the molecule association magnetic field ramp. Molecules are detected for fields below 168.218\,G. This field corresponds to the center of the loss feature shown in Fig.~\ref{fig:FastLoss}, which is marked by the vertical solid line here. The dashed vertical lines mark the $1/e$-width of the loss feature and the horizontal dashed line marks a systematic offset.}
\label{fig:ConvEff}
\end{center}
\end{figure}

We now examine the molecule creation process and properties of the molecules in more detail. First, we determine the magnetic field value of the onset of molecule creation. For this, we perform experiments as the one just described, but we vary the endpoint of the magnetic field ramp, keeping the ramp duration fixed. The frequency of the rf pulse for the separation of free K atoms and LiK molecules and the probe beam frequencies are adapted accordingly. The Li rf pulse was not used in these experiments.

Figure~\ref{fig:ConvEff} shows the lower bound for the molecule fraction $\mathcal{F}$. Imperfect rf pulses lead to a 3\% systematic offset in the data, indicated by the horizontal dashed line \cite{endnote:RFOffset}. It is found that the detected molecule fraction depends strongly on the endpoint of the magnetic field ramp. No molecules are detected down to a final field of 168.217\,G. Only 13\,mG lower the maximum molecule fraction is observed. This magnetic field range corresponds well to the required detuning from resonance for our atom-molecule separation method to work, as discussed above. For lower fields the molecular signal drops again, first steeply down to about 168.19\,G and then much slower. At about 167.5\,G (outside the range of the plotted data) it becomes indiscernible from the background noise. The dependence of the detected molecule fraction on the field may have several reasons. It might be caused by the change in absorption cross section of the molecules with the magnetic field. The slow decrease away from resonance comes from loss of molecules as more time is spent between molecule association and detection.

\begin{figure}[t]
\begin{center}
\includegraphics[width=0.85\columnwidth]{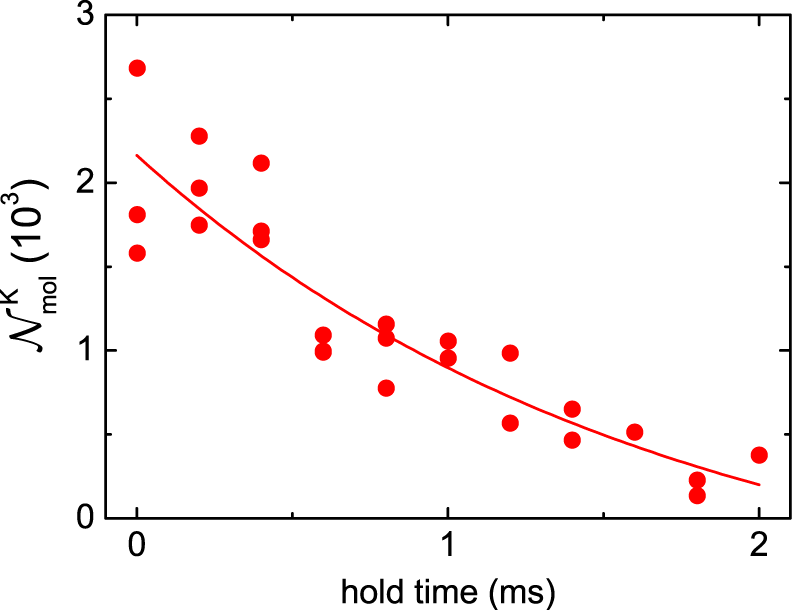}
\caption{Decay of LiK molecules at 168.204\,G. Plotted is the number of molecules $\mathcal{N}^{\rm K}_{\rm mol}$ in dependence on the hold time after the fast magnetic field ramp. The solid line is an exponential fit to the data, yielding a lifetime of 1.7\,ms.}
\label{fig:LiKDecay}
\end{center}
\end{figure}

Within the measurement precision of a few mG the onset of molecule detection coincides with the center of the loss feature from Fig.~\ref{fig:FastLoss}, marked by the solid vertical line in Fig.~\ref{fig:ConvEff}. This observation is in accordance with the standard picture of molecule formation close to a \fbr\ \cite{Kohler2006poc}.

The maximum K molecule conversion efficiency extracted from this data is reached at 168.204\,G and amounts to about 40\%. A different method to determine the K molecule conversion efficiency is to examine the number of free K atoms at a magnetic field just above the onset of molecule production and just below 168.19\,G. Assuming all missing atoms have formed molecules, the molecule conversion efficiency is also determined to be 40\%. The assumption that no molecules are lost is well justified since the time spent in the \fbr\ region during the magnetic field ramp to 168.19\,G (120\,$\mu$s) is short compared to the lifetime of the molecules.

The lifetime of the LiK molecules is determined by holding the sample after molecule creation for a varying time in the ODT at a constant magnetic field of 168.204\,G and measuring the molecule number afterwards. A fit to the decay of the molecule number gives a lifetime of 1.7\,ms; see Fig.~\ref{fig:LiKDecay}. This lifetime does not change if the remaining free Li atoms are removed just after molecule creation by a resonant flash of light, indicating that the dominant loss mechanism does not involve free Li atoms. We did not investigate the effect of unpaired K atoms on the molecule lifetime.

\begin{figure}[t]
\begin{center}
\includegraphics[width=0.89\columnwidth]{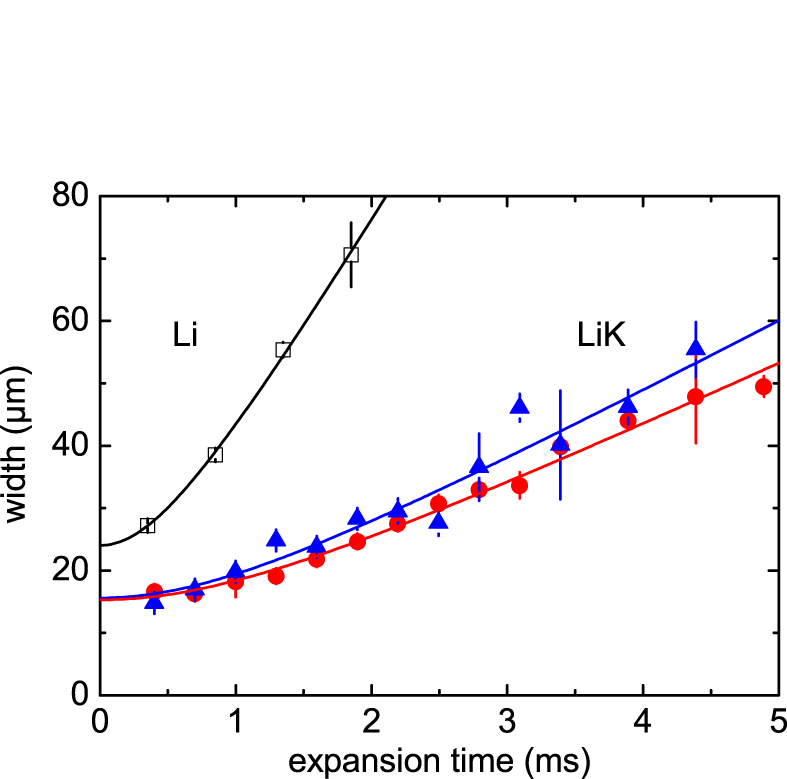}
\caption{Comparison of the free expansion of LiK Feshbach molecules (circles: detection of the bound K atoms, triangles: detection of the bound Li atoms) and unpaired Li atoms (open squares) at 168.196\,G. Shown is the radial $1/\sqrt{e}$-width of Gaussian fits to integrated density profiles.}
\label{fig:LiKMolTOF}
\end{center}
\end{figure}

A striking manifestation of molecule formation can be observed by comparing the expansion behavior of clouds of LiK molecules with the one of clouds of unpaired Li atoms in imaging with Li light. For this comparison, we record the expansion of the molecules and the remaining unpaired Li atoms after a molecule association magnetic field ramp to 168.196\,G; see Fig.~\ref{fig:LiKMolTOF}. We find the average expansion velocity of molecules to be slower by a factor of 3.3, as determined by fits to the expansion. We interpret this difference mainly as a result of the higher mass of the molecules compared to unpaired Li atoms. It corresponds well to the expected velocity ratio of  $v_\mathrm{Li}/v_\mathrm{LiK}=\sqrt{M_\mathrm{LiK}/M_\mathrm{Li}}=\sqrt{46/6}=2.8$ in the approximation of thermal clouds of equal temperature. This observation tells us that Li atoms that remain in state \1\ after the $\pi$-pulse are bound to K atoms.

\subsection{Discussion}
\label{sec:Discussion}

In our experiment, molecule association is achieved in basically the same way as demonstrated in many other cold atom experiments before \cite{Chin2009FBR,Kohler2006poc} and our results agree well with the standard picture of molecule formation close to a \fbr\ \cite{Kohler2006poc}. We observe that molecule association is most efficient in samples of high phase-space density and obtain a maximum molecular conversion efficiency for K of 40\%. This conversion efficiency is typical for experiments employing the Feshbach ramp technique. A Monte Carlo simulation based on the method presented in Ref.~\cite{Hodby2005peo} agrees with our results, giving a conversion efficiency of about 50\% for K.

The lifetime of our molecules is quite short, only 1.7\,ms. Because of this, it would be technically challenging to observe the standard signature of molecule association, which is the reduction of the absorption imaging signal when ramping to the molecular side of the \fbr\ and the recovery of the signal after ramping back. Our rf state separation detection technique, which allows to obtain images of molecules less than 0.1\,ms after molecule association, overcomes this detection problem.

The Li$|1\rangle$K$|1\rangle$ molecule lifetime that we measure is much shorter than typical lifetimes of Li$|1\rangle$K$|3\rangle$ molecules that were measured by the Munich group \cite{Voigt2009uhf}. Presently, we do not know whether the different spin channels and therefore the different Feshbach resonances used for molecule association can explain the different lifetimes. There are also other possible, more technical reasons, which we presently cannot rule out. One possibility, which needs further investigation, is the loss of molecules because of the absorption of photons from the broad spectrum of the multi-mode fiber laser used for the ODT \cite{endnote:MultiFrequencyLaser}.

\section{Conclusion \& Outlook}

We have presented an all-optical evaporative and sympathetic cooling scheme for the preparation of a double degenerate $^6$Li-$^{40}$K Fermi-Fermi mixture. We have also shown the general methodology to prepare the sample close to specific interspecies Feshbach resonances. As a first application, we have demonstrated the formation of Fermi-Fermi heteronuclear molecules and we have examined the molecule association process and some properties of the molecules.

With the basic tools at hand, we are now in the position for the next steps towards our main goal of realizing strongly interacting regimes in the Fermi-Fermi mixture. Since the available Feshbach resonances are quite narrow this requires precise knowledge of the exact resonance position and the magnetic-field dependent elastic and inelastic interaction properties. We are currently inspecting the relatively broad 155-G resonance in the Li|1>-K|3> channel as a promising candidate, for which we experimentally find a width of about 800\,mG \cite{NaikInPreparation}. Strongly interacting conditions generally require a scattering length exceeding the interparticle spacing. Under our typical experimental conditions this would be realized with a magnetic detuning below 10\,mG, which is experimentally feasible.

\begin{acknowledgments}

We want to thank Paul Julienne for fruitful discussions and support on the theoretical understanding of the interspecies scattering properties, Tobias Tiecke for the calculated Feshbach resonance positions displayed in Fig.~\ref{fig:Li2KsWave}, Christoph Kohstall for useful comments on the manuscript, and Clarice Aiello for contributions to the experimental set-up. This work is supported by the Austrian Science Fund (FWF) and the European Science Foundation (ESF) within the EuroQUAM/FerMix project and by the FWF through the SFB FoQuS.

\end{acknowledgments}

\appendix
\section{Magnetic field coils}
\label{apx:magnets}

Three pairs of magnetic field coils are present in the setup: a pair of high-current, large-diameter coils, which we call Feshbach coils, a smaller-diameter pair of high-current coils, which we call curvature coils, and a third, low-current, low inductance pair of coils, which we call fast coils. Normally, the currents in all coils circulate in the same direction. To achieve a quadrupole field configuration for MOT operation, the direction of current in one coil of the Feshbach coil pair and one coil of the curvature coil pair can be reversed using mechanical relays. In the normal configuration, the Feshbach coils are in Helmholtz configuration and give a very homogeneous bias field near the trap center of up to 3000\,G. The curvature coils exhibit a magnetic field curvature, which gives rise to an additional contribution to the trapping potential \cite{Jochim2003bec}. With the current used during evaporation, the curvature coils give a homogeneous bias field of 600\,G and a magnetic field curvature of 27\,G/cm$^2$ along the axial direction of the dipole trap beams (perpendicular to the symmetry axis of the coils). This curvature gives rise to a magnetic confinement corresponding to trap frequencies of 27\,Hz for Li and 10\,Hz for K. When working at bias fields between 150\,G and 170\,G, where the interspecies Feshbach resonances are, the curvature coils provide a magnetic confinement corresponding to 13\,Hz for Li and 5\,Hz for K.

When high magnetic field-stability is needed, we make use of a battery-powered current supply. Since the interspecies \fbr s are very narrow, it is necessary to control the magnetic field with very high precision. Passive stabilization methods, not employing any shielding, lead to a stability of about 10\,mG peak-to-peak over a 50\,Hz cycle. By synchronizing the experimental sequence to line, we achieve a magnetic field stability of a few mG for times on the order of one ms, which is much larger than the typical duration of rf $\pi$-pulses we use for internal state transfer. Magnetic field values are calibrated using rf transitions.

For probing the interspecies resonances we make use of the fast coils, which have Helmholtz configuration. Using these coils we make precise magnetic field ramps of up to 3\,G in about 0.1\,ms. This response time of the magnetic field was characterized by measuring the change in frequency of an atomic RF transition with time after a step change of the current. The response time is not limited by the speed of change of the current through the coil, but by eddy currents.


\begin{thebibliography}{49}
\bibitem[{end({\natexlab{a}})}]{endnote:HighFieldImaging}
\bibinfo{note}{We use high-field imaging at 1190\,G where the scattering length
  between Li$|1\rangle$ and $|2\rangle$ is small and negative. The Li cloud can
  thus be approximated as a weakly-interacting Fermi gas which in the
  degenerate regime allows for easier interpretation of the profiles obtained
  by absorption images after time-of-flight.}
\bibitem[{end({\natexlab{b}})}]{endnote:RFTransitions}
\bibinfo{note}{The rf transition frequencies for Li and K are 39\,MHz and
  69\,MHz, respectively. The $\pi$-pulses are approximately rectangular and
  have a total duration of 43\,$\mu$s (333\,$\mu$s) for K (Li). The
  duration of the Li pulse is much longer than that of the K pulse, because
  less rf power is available. The full width of the dominant central peak in
  the Fourier spectrum is 46\,kHz for K and 6\,kHz for Li. We can achieve
  transfer efficiencies higher than 99\,\%.}
\bibitem[{end({\natexlab{c}})}]{endnote:MoleculeImaging}
\bibinfo{note}{A slight off-resonant contribution of atoms in the K$|2\rangle$
  state is visible on K$|1\rangle$ images. Since this contribution has a much
  larger spatial extent than the cloud of molecules, it is possible to correct
  for it when extracting data from the image. We also account for unpaired
  atoms not transferred by the rf pulse by independently measuring the
  $\pi$-pulse transfer efficiency.}
\bibitem[{end({\natexlab{d}})}]{endnote:unpairedAtomImaging}
\bibinfo{note}{Li$|2\rangle$ atoms are directly imaged using a probe beam of
  appropriate frequency. Since we do not have a probe beam with the correct
  frequency to image K$|2\rangle$ atoms, we first transfer those atoms back to
  state $|1\rangle$ using another $\pi$-pulse and then take the image using the
  same K$|1\rangle$ probe beam as before. Atoms imaged on the first set of
  pictures have been accelerated and heated by the probe beam flash and have
  dispersed enough at the time the second image is taken to not influence it.}
\bibitem[{end({\natexlab{e}})}]{endnote:RFOffset}
\bibinfo{note}{The offset in dependence of the magnetic field originates from
  imperfect frequency adjustment of the K $\pi$-pulse and was determined by
  experiments during which molecule formation was inhibited by removing
  Li$|1\rangle$ from the sample with a short flash of resonant light before the
  molecule association magnetic field ramp. The offset is 1\,\% at 168.9\,G and
  rises to 10\,\% 1.5\,G away from that position.}
\bibitem[{end({\natexlab{f}})}]{endnote:MultiFrequencyLaser}
\bibinfo{note}{A strong influence of broadband laser light on the lifetime of
  molecules was observed in experiments on cesium Feshbach molecules in
  Innsbruck. The lifetime substantially increased when the trapping light was
  provided by a single-mode infrared laser.}
\end{thebibliography}

\end{document}